\documentclass[aps]{revtex4}

\usepackage{amsmath,amssymb,amsfonts}
\usepackage{graphicx}
\usepackage{color}
\allowdisplaybreaks
\def\be{\begin{equation}}
\def\ee{\end{equation}}
\def\bea{\begin{eqnarray}}
\def\eea{\end{eqnarray}}
\def\ba{\bea}
\def\ea{\eea}
\def\bm{\bibitem}
\def\rw{\rightarrow}

\def\del{\partial}

\def\ra{\rangle}
\def\la{\langle}

\def\no{\nonumber}
\def\negs{\!\!\!\!\!\!\!\!}
 
\def\bet{\beta}

\def\de{\delta}
\def\ep{\epsilon}

\def\sg{\sigma}
\def\th{\theta}
\def\om{\omega}
\def\Gm{\Gamma}
\def\De{\Delta}
\def\Lm{\Lambda}

\def\tht{\th}
\def\mn{{\mu\nu}}
     
\def\vk{\boldsymbol{k}}    
\def\vq{\boldsymbol{q}} 
\def\vl{\boldsymbol{l}} 
\def\vx{\boldsymbol{x}}     
    
\def\vph{\vec\phi}

\def\bld{\boldsymbol}
\def\bD{\bld{D}}

\def\bU{\boldsymbol{U}}

\def\bL{\boldsymbol{\Lambda}}

\def\ov{\overline} 

\def\oD{\overline{D}}

\def\cm {{\cal M}}
\def\cl {{\cal L}}
\def\omp {\om_\pi}
\def\omh {\om_h}

\begin{document}

\setcounter{page}{1}

\title{Viscous coefficients of a hot pion gas} 
\author{Sourav \surname{Sarkar}}
\affiliation{Theoretical Physics Division, Variable Energy Cyclotron
Centre, 1/AF, Bidhannagar, Kolkata, 700064, India}


\begin{abstract}
The steps 
essentially involved in the evaluation of transport coefficients 
in linear response theory using Kubo formulas are 
to relate the defining {\em retarded} correlation function to the corresponding 
{\em time-ordered} one and to evaluate the latter in the conventional
perturbation expansion. Here we evaluate the viscosities of a pion gas
carrying out both the steps in the {\em real time}
formulation. We also obtain the viscous coefficients by solving the relativistic transport
equation in the Chapman-Enskog approximation to
leading order. An in-medium $\pi\pi$ cross-section in used in which spectral
modifications are introduced in the propagator of the exchanged $\rho$. 
\end{abstract}


\maketitle

\section{Introduction}
\setcounter{equation}{0}
\renewcommand{\theequation}{1.\arabic{equation}}

One of the most interesting results from experiments at the Relativistic Heavy
Ion Collider (RHIC) is the surprisingly large magnitude of the elliptic flow of
the emitted hadrons. Viscous hydrodynamic simulations of heavy ion collisions
require a rather small value of $\eta/s$, $\eta$ being the coefficient of shear
viscosity and $s$ the entropy density, for the theoretical interpretation of this
large collective flow. The value being close to $1/4\pi$, the quantum lower bound for
this quantity~\cite{KSS}, matter produced in these collisions is believed to be
almost a perfect fluid~\cite{Csernai}.

This finding has led to widespread interest in the study of non-equilibrium
dynamics, especially in the microscopic evaluation of the transport coefficients of
both partonic as well as hadronic forms of strongly interacting matter. In the
literature one comes across basically two approaches that have been used to
determine these quantities. One is the kinetic theory method in which the
non-equilibrium distribution function which appears in the transport equation is
expanded in terms of the gradients of the flow velocity field. The coefficients
of this expansion which are related to the transport coefficients are then
perturbatively determined using various approximation methods. The other approach is based on response theory
in which the non-equilibrium transport coefficients are related by Kubo formulas
to equilibrium correlation functions. They are then perturbatively evaluated 
using the techniques of thermal field theory. Alternatively, the Kubo formulas can be
directly evaluated on the lattice~\cite{Meyer} or in transport cascade
simulations~\cite{Demir} to obtain the transport coefficients.

Thermal quantum field theory has been formulated in the imaginary as well
as real time \cite{Matsubara,Mills,Umezawa,Niemi,Kobes}. For time independent 
quantities such as the partition function, the imaginary time formulation is
well-suited and stands as the only simple method of calculation. However,
for time dependent quantities like two-point correlation functions, the use of 
this formulation requires a continuation to imaginary time and possibly back to
real time at the end. On the other hand, the real time formulation provides
a convenient framework to calculate such quantities, without requiring any
such continuation at all.

A difficulty with the real time formulation is, however, that all two-point 
functions take the form of $2\times 2$ matrices. But this difficulty is only 
apparent: Such matrices are always diagonalisable and it is the $11$- component of 
the diagonalised matrix that plays the role of the single function in the
imaginary time formulation. It is only in the calculation of this $11$-component 
to higher order in perturbation that the matrix structure appears in a non-trivial 
way.

In the literature transport coefficients are evaluated using the imaginary time 
formulation \cite{Hosoya,Lang,Jeon}. Such a coefficient is defined by the 
{\em retarded} correlation function of the components of the energy-momentum tensor. 
As the conventional perturbation theory applies only to {\em time-ordered}
correlation functions, it is first necessary to relate the two types of
correlation functions using the K\"{a}llen-Lehmann spectral representation
~\cite{Kallen,Lehmann,Fetter,MS}. We find this relation directly in real time 
formulation. The time-ordered correlation function is then calculated also in the 
covariant real time perturbative framework to finally obtain the the viscosity
coefficients of a pion gas.

We also calculate the viscous coefficients in a kinetic theory framework
by solving the transport equation in the Chapman-Enskog approximation to leading
order. This approach being computationally more
efficient~\cite{Jeon},
has been mostly used in the literature to obtain the viscous coefficients. 
The $\pi\pi$ cross-section is a crucial
dynamical input in these calculations. Scattering amplitudes evaluated using chiral perturbation
theory~\cite{Weinberg1,Gasser} to lowest order have been used 
in~\cite{Santalla,Chen}
and unitarization improved estimates of the amplitudes were used 
in~\cite{Dobado3} to evaluate the shear viscosity. Phenomenological
scattering cross-section using experimental phase shifts have been
used in~\cite{Prakash,Chen,Davesne,Itakura} in view of the fact that  
the $\pi\pi$ cross-section estimated from lowest order chiral perturbation theory
is known to deviate from the experimental data beyond centre of mass energy of 500 MeV
primarily due to the $\rho$ pole which dominates the cross-section in the
energy region between 500-1000 MeV. All these approaches have used a
vacuum cross section. To construct an in-medium cross-section 
we employ an effective Lagrangian approach which incorporates
$\rho$ and $\sigma$ meson exchange in $\pi\pi$ scattering. Medium effects are
then introduced in the $\rho$ propagator through one-loop self-energy
diagrams~\cite{Sukanya1}.

In Sec.~II we derive the spectral representations for the retarded  and time-ordered 
correlation functions in the real time version of thermal field theory. 
We also review the formulation of the non-equilibrium density operator and 
obtain the expressions
for the viscosities in terms of equilibrium (retarded) two-point functions.
The time-ordered function is then calculated to lowest order with complete 
propagators in the equilibrium theory.
In Sec.~III we briefly recapitulate the expressions for the viscosities
obtained by solving the Uehling-Uhlenbeck transport equation in the 
kinetic theory framework. We then evaluate the $\pi\pi$ cross-section in the 
medium briefly discussing the one-loop $\rho$ self-energy due to
$\pi h(h=\pi,\om,h_1,a_1)$ loops evaluated in the real-time formulation discussed
above. We end with a summary in
Sec.~IV.

\section{Viscous coefficients in the linear response theory}
\setcounter{equation}{0}
\renewcommand{\theequation}{2.\arabic{equation}}
\subsection{Real-time formulation}
\begin{figure}
\centerline{\includegraphics[scale=0.7]{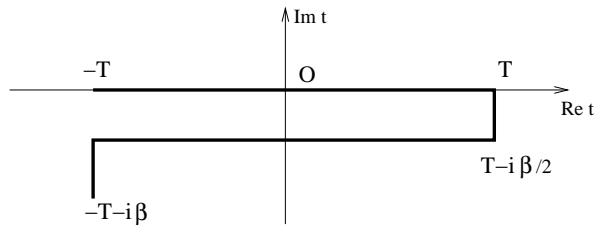}}
\caption{The contour $C$ in the complex time plane used here for the real
time formulation.}
\end{figure}

In this section we review the real time formulation of equilibrium 
thermal field theory leading to the
spectral representations of bosonic two-point functions \cite{MS}. 
This formulation 
begins with a comparison between the time evolution operator
$e^{-iH(t_2-t_1)}$ of quantum theory and the Boltzmann weight $e^{-\bet H}
=e^{-iH(\tau-i\bet-\tau)}$ of statistical physics, where we introduce $\tau$ as a 
complex variable. Thus while for the time evolution operator, the times $t_1$ and
$t_2$ $~(t_2 > t_1)$ are any two points on the real line, the Boltzmann
weight involves a path from $\tau$ to $\tau-i\bet$ in the complex time
plane. Setting this $\tau =-T$, where $T$ is real, positive and large, we
can get the contour $C$ shown in Fig.~1, lying within the region of analyticity
in this plane and accommodating real time correlation functions \cite{Mills,Niemi}.

Let a general bosonic interacting field in the Heisenberg representation be
denoted by $\Phi_l(x)$, whose subscript $l$ collects the index (or indices)
denoting the field component and derivatives acting on it. Although we shall
call its two-point function as propagator, $\Phi_l(x)$ can be an elementary 
field or a composite local operator. (If $\Phi_l(x)$ denotes the pion field, it 
will, of course, not have any index).

The thermal expectation value of the product $\Phi_l(x)\Phi^\dag_{l'}(x')$ may
be expressed as
\be
\la \Phi_l(x)\Phi^\dag_{l'}(x')\ra=\frac{1}{Z}\sum_{m,n}e^{-\beta E_m} \la
m|\Phi_l(x)|n\ra\la n|\Phi^\dag_{l'}(x')|m\ra\,,
\label{expctvalue}
\ee
where $\la O\ra$ 
for any operator $O$ denotes {\em equilibrium} ensemble average;
\be
\la O\ra = {\rm Tr}(e^{-\beta H} O)/Z\,, ~~~~~~~~Z=Tre^{-\beta H}\,.
\ee
Note that we have two sums in (\ref{expctvalue}), one to evaluate the trace and the other
to separate the field operators. 
They run over a complete set of states,
which we choose as eigenstates $|m\ra$ of four-momentum $P_\mu$. Using 
translational invariance of the field operator,
\be
\Phi_l(x)=e^{iP\cdot x}\Phi_l(0)e^{-iP\cdot x}\,,
\ee
we get
\be
\la\Phi_l(x)\Phi^\dag_{l'}(x')\ra=\frac{1}{Z}\sum_{m,n}e^{-\beta E_m}\,e^{i(k_m-k_n)\cdot (x-x')}
\la m|\Phi_l(0)|n\ra\la n|\Phi^\dag_{l'}(0)|m\ra\,.
\ee
Its spatial Fourier transform is
\bea
&&\negs \int d^3x\,e^{-i\vk\cdot(\vx-\vx')}\la\Phi_l(x)\Phi^\dag_{l'}(x')\ra\no\\
&&\negs =\frac{(2\pi)^3}{Z} \sum_{m,n}e^{-\beta E_m}\,e^{i(E_m-E_n)(\tau-\tau')}
\de^3(\vk_m-\vk_n+\vk)\la m|\Phi_l(0)|n\ra\la n|\Phi^\dag_{l'}(0)|m\ra\,,
\label{spatFT}
\eea
where the times $\tau,\, \tau'$ are on the contour $C$. We now insert unity on the 
left of eq.~(\ref{spatFT}) in the form
\[1=\int_{-\infty}^\infty dk_0' \de(E_m-E_n+k_0')\,.\]
(We reserve $k_0$ for the variable conjugate to the real time.) Then it may be 
written as 
\be
\int d^3x \,e^{-i\vk\cdot(\vx-\vx')}\la\Phi_l(x)\Phi^\dag_{l'}(x')\ra
=\int\frac{dk_0'}{2\pi}e^{-ik_0'(\tau-\tau')}M^+_{ll'}(k_0',\vk)\,,
\label{ft1}
\ee
where the spectral function $M^+$ is given by $[k'_\mu=(k_0',\vk)]$
\be
M^+_{ll'}(k')=\frac{(2\pi)^4}{Z}\sum_{m,n}e^{-\beta E_m}\,\de^4(k_m-k_n+k')
\la m|\Phi_l(0)|n\ra\la n|\Phi^\dag_{l'}(0)|m\ra\,.
\label{mplus}
\ee

In just the same way, we can work out the Fourier transform of 
$\la\Phi^\dag_{l'}(x')\Phi_l(x)\ra$ 
\be
\int d^3x \,e^{-i\vk\cdot(\vx-\vx')}\la\Phi^\dag_{l'}(x')\Phi_l(x)\ra
=\int\frac{dk_0'}{2\pi}e^{-ik_0'(\tau-\tau')}M^-_{ll'}(k_0',\vk)\,,
\label{ft2}
\ee
with a second spectral function $M^-$ is given by
\be
M^-_{ll'}(k')=\frac{(2\pi)^4}{Z}\sum_{m,n}e^{-\beta E_m}\,\de^4(k_n-k_m+k')
\la m|\Phi^\dag_{l'}(0)|n\ra\la n|\Phi_l(0)|m\ra\,.
\label{mminus}
\ee
The two spectral functions are related by the KMS relation \cite{Kubo,Martin}
\be
M^+_{ll'}(k)=e^{\beta k_0}M_{ll'}^-(k)\,,
\label{KMS}
\ee
in momentum space, which may be obtained simply by interchanging the dummy indices 
$m,n$ in one of $M^\pm_{ll'}(k)$ and using the energy conserving $\de$-function.

We next introduce the {\it difference} of the two spectral functions,
\be
\rho_{ll'}(k) \equiv M_{ll'}^+(k)-M_{ll'}^-(k)\,,
\label{diff}
\ee
and solve this identity and the KMS relation (\ref{KMS}) for $M^\pm_{ll'}(k)$, 
\be
M^+_{ll'} (k)=\{1+f(k_0)\}\rho_{ll'} (k)\,, ~~~~ M^-_{ll'}(k)=f(k_0)\rho_{ll'}(k)\,,
\label{KMS1}
\ee
where $f(k_0)$ is the distribution-like function
\be
f(k_0)=\frac{1}{e^{\beta k_0}-1}\,,~~~~~~~~-\infty <k_0 < \infty\,.
\ee
In terms of the true distribution function
\be 
n(|k_0|)=\frac{1}{e^{\beta |k_0|}-1}\,,
\ee
it may be expressed as
\bea
f(k_0)&=& f(k_0)\{\tht (k_0)+\tht(-k_0)\}\no\\
&=&  n\ep(k_0)-\tht (-k_0)\,.
\label{ff1}
\eea

With the above ingredients, we can build the spectral representations for the two 
types of thermal propagators. First consider the {\em time-ordered} one,
\bea
-iD_{ll'}(x,x')&=&\la T_c \Phi_l(x) \Phi^\dag_{l'}(x')\ra \no\\
&=&\tht_c(\tau-\tau')\la \Phi_l(x) \Phi^\dag_{l'}(x')\ra+\tht_c(\tau'-\tau)
\la \Phi^\dag_{l'}(x') \Phi_{l}(x)\ra\,.
\eea
Using eqs.~(\ref{ft1},\ref{ft2},\ref{KMS1}), we see that its spatial Fourier transform is given by
 \cite{Mills}
\be
D_{ll'}(\tau-\tau',\vk)=i\int_{-\infty}^\infty\frac{dk_0'}{2\pi}\rho_{ll'}(k_0',\vk)
e^{-ik_0'(\tau-\tau')}\{\tht_c(\tau-\tau')+f(k_0')\}\,,
\ee

As $T\rw \infty$, the contour of Fig.~1 simplifies, reducing essentially to
two parallel lines, one the real axis and the other shifted by $-i\bet/2$,
points on which will be denoted respectively by subscripts 1 and $2$, so
that $\tau_1=t,\, \tau_2=t-i\bet/2$ \cite{Niemi}. The propagator then consists of 
four pieces, which may be put in the form of a $2\times 2$ matrix.
The contour ordered $\th's$ may now be converted to the usual time ordered
ones. If $\tau,\tau'$ are both on line $1$ (the real axis), the $\tau$ and
$t$ orderings coincide, $\th_c(\tau_1-\tau'_1)=\th(t-t')$. If they are on
two different lines, the $\tau$ ordering is definite,
$\th_c(\tau_1-\tau'_2)=0,\, \th_c(\tau_2-\tau'_1)=1$. Finally if they are
both on line $2$, the two orderings are opposite,
$\th_c(\tau_2-\tau'_2)=\th(t'-t)$.

Back to real time, we can work out the usual temporal Fourier transform of
the components of the matrix to get
\be   
\bD_{ll'} (k_0,\vk)=\int_{-\infty}^{\infty}\frac{dk_0'}{2\pi}\rho_{ll'}(k_0',\vk)
\bL (k_0',k_0)\,,  
\label{d11matrix}
\ee
where the elements of the matrix $\bL$ are given by \cite{MS}
\bea
&& \Lm_{11}=-\Lm_{22}^* =\frac{1}{k_0'-k_0-i\eta}+2\pi if(k_0')\de(k_0'-k_0)\,,
\no\\
&& \Lm_{12}=\Lm_{21}=2\pi ie^{\beta k_0'/2}f(k_0')\de(k_0'-k_0)\,.
\label{Lm1}
\eea
Using relation (\ref{ff1}), we may rewrite (\ref{Lm1}) in terms of $n$, 
\bea
&& \Lm_{11}=-\Lm_{22}^* =\frac{1}{k_0'-k_0-i\eta\ep(k_0)}+2\pi in\ep(k_0)
\de(k_0'-k_0)\,,\no\\
&& \Lm_{12}=\Lm_{21}=2\pi i\sqrt{n(1+n)}\ep(k_0)\de(k_0'-k_0)\,.
\label{Lm2}
\eea

The matrix $\bL$ and hence the propagator $\bD_{ll'}$ can be diagonalised
to give 
\be
\bD_{ll'}(k_0,\vk)=\bU 
\left(\begin{array}{cc}\oD_{ll'} & 0\\0 &
-\oD_{ll'}^{\,*}\end{array}\right)\bU\,,
\label{diag}
\ee
where $\oD_{ll'}$ and $\bU$ are given by
\be
\oD_{ll'} (k_0,\vk)=\int_{-\infty}^{\infty}
\frac{dk_0'}{2\pi}\frac{\rho_{ll'}(k_0',\vk)}{k_0'-k_0-i\eta\ep(k_0)}\,,~~~~~~
\bU = \left( \begin{array}{cc} \sqrt{1+n} & \sqrt{n}\\\sqrt{n} &
\sqrt{1+n}\end{array}\right)\,.
\label{uuu}
\ee 
Eq.~(\ref{diag}) shows that $\oD$ can be obtained from any of the elements of the
matrix $\bD$, say $D_{11}$. Omitting the indices $ll'$, we get
\be
{\rm Re}\oD={\rm Re D}_{11}\,,~~~~{\rm Im}\oD=\tanh(\bet|k_0|/2){\rm Im D}_{11}\,.
\label{tanh}
\ee

Looking back at the spectral functions $M^{\pm}_{ll'}$  defined by
(\ref{mplus}, \ref{mminus}), we can 
express them as usual four-dimensional Fourier transforms of ensemble average of 
the operator products, so that $\rho_{ll'}$ is the Fourier transform of that of the 
commutator,
\be
\rho_{ll'}(k_0,\vk)=\int d^4y e^{ik\cdot (y-y')}\la
[\Phi_l(y),\Phi_{l'}(y')]\ra\,,
\ee
where the time components of $y$ and $y'$ are on the real axis in the
$\tau$-plane. Taking the spectral function for the free scalar field,
\be
\rho_0=2\pi\ep(k_0)\de (k^2-m^2)\,,
\label{freesp}
\ee
we see that $\oD$ becomes the free propagator, $\oD (k_0,\vk)=-1/(k^2-m^2)$.

We next consider the {\em retarded} thermal propagator
\be
D^R_{ll'}(x,x')=i\th_c(\tau-\tau')\la[\Phi_l(\vx,\tau),\Phi_{l'}(\vx',\tau')]\ra\,,
\ee
where again $\tau\,,\tau'$ are on the contour $C$ (Fig.~1). Noting
eqs.~(\ref{ft1},\ref{ft2},\ref{diff}) the three dimensional Fourier transform may immediately
be written as 
\be
D^R_{ll'}(\tau-\tau', \vk)=i\th_c(\tau-\tau')\int_{-\infty}^\infty
\frac{dk_0'}{2\pi}e^{-ik_0'(\tau-\tau')}\rho_{ll'} (k_0',\vk)\,.
\ee
As before we isolate the different components with real times and take 
the Fourier transform with respect to real time. Thus for the $11$-component we 
simply have
\be
D^R_{ll'}(t-t',\vk)_{11}=i\th (t-t')\int_{-\infty}^\infty
\frac{dk_0'}{2\pi}e^{-ik_0'(t-t')}\rho_{ll'} (k_0',\vk)\,,
\ee
whose temporal Fourier transform gives 
\be
D^R_{ll'}(k_0,\vk)_{11}=\int_{-\infty}^\infty             
\frac{dk_0'}{2\pi}\frac{\rho_{ll'} (k_0',\vk)}{k_0'-k_0-i\eta}\,.      
\label{dr}
\ee
This $11$-component suffices for us, but we also display the complete matrix,
\be
\bD^R_{ll'}(k_0,\vk)= \left(\begin{array}{cc} D^R_{ll'}(k)_{11} & 0\\
\rho_{ll'}(k) \{\sqrt{\frac{n}{n+1}}\th(k_0)+\sqrt{\frac{n+1}{n}}\th(-k_0)\} &
-D^{R*}_{ll'}(k)_{11}\end{array}\right)\,.
\ee
Though we deal with matrices in real time formulation, it is the $11$-component
that is physical. Eqs.~(\ref{uuu}) and (\ref{dr}) then show that we can continue the
{\em time-ordered} two-point function into the {\em retarded} one by simply 
changing the $i\ep$ prescription,
\be
D^R_{ll'}(k_0+i\eta,\vk)_{11}=\ov D_{ll'}(k_0+i\eta\ep(q_0)\to k_0+i\eta,\vk)\,.
\ee 
The point to note here is that for the time-ordered propagator, it is the
{\em diagonalised} matrix and not the matrix itself, whose $11$-component can be
continued in a simple way.

\subsection{Transport coefficients}

We now use the linear response approach to arrive at expressions of the
transport coefficients as integrals of retarded Green's functions over space.
We follow the method proposed by Zubarev~\cite{Zubarev}, which is excellently reviewed
in~\cite{Hosoya}. Here the system is supposed to be in
the hydrodynamical stage where the mean free time of the constituent particles
is much shorter than the relaxation time of the whole system under consideration.
Thus local equilibrium will be attained quickly, while global
equilibrium will be approaching gradually. Since the system is assumed to be not far
from equilibrium, we may retain only linear terms in space-time gradients of
thermodynamical parameters, like temperature and velocity fields. We assume 
the energy-momentum of the system to be conserved,
\be
\del_\mu T^\mn(\vx,t)=0~.
\label{dtmn}
\ee  
The non-equilibrium density matrix operator is constructed in the Heisenberg
picture, where it is independent of time,
\be
\frac{d\rho}{dt}=0~.
\ee
Following Zubarev, we construct the operator $B(\vx,t)$,
\be
B(\vx,t)=\ep\int_{-\infty}^t dt_1
e^{\ep(t_1-t)}F^\nu(\vx,t_1)T_{0\nu}(\vx,t_1)~,~~~~(\ep\to 0^+)
\label{Bxt}
\ee
where $F^\nu(\vx,t)=\beta(\vx,t)u^\nu(\vx,t)$. Here $\beta(\vx,t)$ is a Lorentz
invariant quantity defining the local temperature and $u^\nu(\vx,t)$ is the
four-velocity field of the fluid,
\be
u^\nu(\vx,t)u_\nu(\vx,t)=1~.
\ee
The construction (\ref{Bxt}), which smooths out the oscillating terms resemble the one
used in the formal theory of scattering~\cite{Zubarev,Gell-Mann} and selects
out the retarded solution.

The expression (\ref{Bxt}) is actually independent of $t$; the time derivative is
\be
\frac{d}{dt}B(\vx,t)=\ep F^\nu(\vx,t)T_{0\nu}(\vx,t)-\ep^2\int_{-\infty}^t
dt_1 e^{\ep(t_1-t)}F^\nu(\vx,t_1)T_{0\nu}(\vx,t_1)~.
\label{dBdt}
\ee
As $T_{0\nu}$ and $F^\nu$ are finite, the right hand side of (\ref{dBdt}) goes to zero as $\ep\to
0$. Also integrating (\ref{Bxt}) by parts, we get
\be
B(\vx,t)=F^\nu(\vx,t)T_{0\nu}(\vx,t)-\int_{-\infty}^tdt_1 e^{\ep(t_1-t)}
\left(F^\nu\frac{dT_{0\nu}}{dt}+\frac{dF^\nu}{dt}T_{0\nu}\right)~.
\label{Bxt2}
\ee
We now consider the space integral of (\ref{Bxt2}). Using the energy-momentum
conservation rule (\ref{dtmn}), we integrate the second term in (\ref{Bxt2}) by parts and
neglect the surface integrals to get
\ba
\int d^3x B(\vx,t)&=&\int d^3x F^\nu(\vx,t)T_{0\nu}(\vx,t)\nonumber\\
&-&\int d^3x\int_{-\infty}^tdt_1 e^{\ep(t_1-t)}T_{\mn}(\vx,t_1)\del^\mu F^\nu(\vx,t_1)
\nonumber\\
&\equiv &A-B
\label{aminusb}
\ea
where we have abbreviated the first and second terms by $A$ and $-B$  
respectively. Then the non-equilibrium statistical density matrix is given by
\be
\rho=e^{-A+B}/{\rm Tr}e^{-A+B}~.
\ee
The first term $A$ in eq.~(\ref{aminusb}) characterises local equilibrium,
\be
\rho_0=e^{-A}/{\rm Tr}e^{-A}
\ee
while the second term $B$ including the thermodynamical force $\del^\mu F^\nu$
describes deviation from equilibrium.

In order to expand $\rho$ in a series in  $B$ we define the function
\be
Q(\tau)=e^{-(1-\tau)A}e^{\tau(-A+B)}
\ee
such that the boundary conditions at $\tau=0$ and $\tau=1$ correspond to the
equilibrium and non-equilibrium density matrices,
\be
Q(\tau=0)=e^{-A}=\rho_0~,~~~~Q(\tau=1)=e^{-A+B}=\rho~.
\ee
We then differentiate $Q(\tau)$ w.r.t $\tau$ to get
\be
\frac{dQ(\tau)}{d\tau}=e^{-(1-\tau)A}Be^{(1-\tau)A}Q(\tau)
\ee
which can be integrated to give
\be
Q(\tau)=Q(0)+\int_0^\tau d\tau'e^{-(1-\tau')A}Be^{(1-\tau')A}Q(\tau')~.
\ee
It can be solved iteratively. Keeping up to the first order term (linear response) and setting
$\tau=1$ we get the required result
\be
\rho=\rho_0\left[1+\int_0^1e^{-\tau A}Be^{\tau A}\right]~.
\ee

Applying this formula to the energy-momentum tensor, we get its response to the
thermodynamical forces as~\cite{Hosoya}
\ba
\la T_{\mn}(\vx,t)\ra&=&\la T_{\mn}(\vx,t)\ra_0\nonumber\\
&+&\int
d^3x'\int_{-\infty}^tdt'e^{\ep(t_1-t)}(T_{\mn}(\vx,t),T_{\rho\sg}(\vx',t'))
\del^\rho F^\sg(\vx',t')
\label{tmn}
\ea
where
\ba
(T_{\mn}(\vx,t),T_{\rho\sg}(\vx',t'))&=&\int_0^1 d\tau \left\{\la T_{\mn}(\vx,t)
e^{-\tau A}T_{\rho\sg}(\vx',t')e^{\tau A}\ra_0\right.\nonumber\\
&&\left.-\la T_{\mn}(\vx,t)\ra_0\la T_{\rho\sg}(\vx',t')\ra_0\right\}
\ea
is the correlation function to be evaluated. As the correlation is assumed 
to vanish as $t'\to -\infty$, it can be put in terms of the conventional retarded Green's
function. Omitting indices it is
\be
(T(\vx,t),T(\vx',t'))=\frac{1}{\beta}\int_{-\infty}^tdt'\la
T(\vx,t),T(\vx',t')\ra_{ret}
\ee
with
\be
\la T(\vx,t),T(\vx',t')\ra_{ret}\equiv
i\theta(t-t')\la[T(\vx,t),T(\vx',t')]\ra_0
\ee

We now use eq.~(\ref{tmn}) to obtain the expectation value of the viscous-shear
stress part of the non-equilibrium energy momentum tensor which is given by
\be
T^\mn=T^{\mn\,(0)}+\pi^\mn+(P^\mu u^\nu+P^\nu u^\mu)
\ee
where $T^{\mn\,(0)}=(\ep+p)u^\mu u^\nu-g^\mn p$ is the equilibrium part, $\pi^\mn$ is the viscous-shear
stress tensor and $P^\mu$ is the heat current. Also, with a view to separate
scalar, vector and tensor processes the quantity
$T_{\rho\sg}\del^\rho F^\sg$ in (\ref{aminusb}) is expanded as
\be
T_{\rho\sg}\del^\rho F^\sg=\beta\pi_{\rho\sg}\del^\rho u^\sg+\beta
P_\rho(\beta\del^\rho\beta+u\cdot\del u^\rho)-\beta \widetilde{p}\del\cdot u
\ee
with $\widetilde{p}=p-c_s^2\ep$, $c_s$ being the sound velocity. Using now the fact that
the correlation function between operators of
different ranks vanish in an isotropic medium, one can write from (\ref{tmn})
\ba
\la \pi_{\mn}(\vx,t)\ra&=&\la \pi_{\mn}(\vx,t)\ra_0\nonumber\\
&+&\int
d^3x'\int_{-\infty}^tdt'e^{\ep(t_1-t)}(\pi_{\mn}(\vx,t),\pi_{\rho\sg}(\vx',t'))
\beta(\vx',t')\del^\rho u^\sg(\vx',t')
\label{pimn}
\ea
with $\la\pi_\mn(\vx,t)\ra_0=0$. Following Hosoya~\cite{Hosoya}
we write the correlation function as
\be
(\pi_{\mn},\pi_{\rho\sg})=\frac{1}{10}[\De_{\mu\rho}\De_{\nu\sg}+
\De_{\mu\sg}\De_{\nu\rho}-\frac{2}{3}\De_\mn\De_{\rho\sg}]
(\pi^{\alpha\beta},\pi_{\alpha\beta})
\ee
where $\De_\mn=g_\mn-u_\mu u_\nu$.
Assuming now that changes
in the thermodynamic forces are small over the correlation length of the two
point function,  the factor $\beta\del^\rho u^\sg$ can be
taken out of the integral giving finally
\be
\la\pi^\mn(\vx,t)\ra=\eta[\De^\mu_\rho\De^\nu_\sg(\del^\rho u^\sg+\del^\sg
u^\rho)-\frac{2}{3}\De^\mn\De_{\rho\sg}\del^\rho u^\sg]
\ee
where
\ba
\eta&=&\frac{1}{10}\int_{-\infty}^0 dt_1 e^{\ep(
t_1-t)}\int_{-\infty}^{t_1}dt'\,\int d^3x'\, 
\la \pi^{\alpha\beta}(\vx,t),\pi_{\alpha\beta}(\vx',t')\ra_{ret}\nonumber\\
&=&\frac{1}{10}\int_{-\infty}^0 dt_1 e^{\ep
t_1}\int_{-\infty}^{t_1}dt'\,i\int d^3x'\, \th (-t')
\la [\pi^{\alpha\beta}(0,0),\pi_{\alpha\beta}(\vx',t')]\ra_0~;~~~~~~(\vx\to 0,t\to 0).
\label{eta}
\ea

Again, starting with the pressure $p(\vx,t)$ on the l.h.s of eq.~(\ref{tmn}) and
following the steps as described above we obtain
\be
\la p(\vx,t) \ra=\la p(\vx,t) \ra_0-\zeta\del_\rho u^\rho(\vx,t)~.
\ee
where the bulk viscosity $\zeta$ is given
in terms of a retarded correlation function by
\be
\zeta=\int_{-\infty}^0 dt_1 e^{\ep(
t_1-t)}\int_{-\infty}^{t_1}dt'\,\int d^3x'\, 
\la \widetilde{p}(\vx,t),\widetilde p(\vx',t')\ra_{ret}~.
\ee
Here $\widetilde p(\vx,t)=p(\vx,t)-c_s^2\ep(\vx,t)$ with 
$\ep(\vx,t)$ the energy density and $c_s^2=\la\del p/\del \ep\ra_0$. 

Recall that $\la \cdots\ra_0$ 
denotes {\em equilibrium} ensemble average.
From now on we shall drop the subscript '0' on the correlation functions.

\subsection{Perturbative evaluation}

\begin{figure}
\centerline{\includegraphics[scale=0.7]{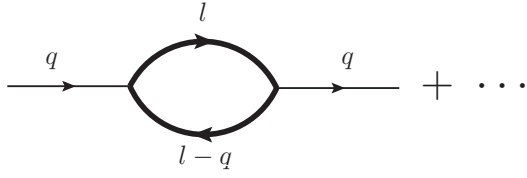}}
\caption{The first term in the so-called skeleton expansion of the two-point function.
 Heavy lines denote full propagators.}
\end{figure}

Clearly the spectral forms and their inter-relations derived in Sec.~IIA hold also
for the two-point function appearing in eq.~(\ref{eta}) for the shear viscosity. We
begin with four-dimensional Fourier transforms. To calculate the $11$-element of 
the the retarded two-point function
\be
\Pi_{11}^R(q)=i\int d^4x e^{iq(x-x')}\th(t-t')\la 
[\pi_{\alpha\beta}(\vx,t),\pi^{\alpha\beta}(\vx',t')]\ra\,,
\label{pi11r1}
\ee
we consider the corresponding time-ordered one,
\be
\Pi_{11}(q)=i\int d^4x e^{iq(x-x')}\la T
\pi_{\alpha\beta}(\vx,t)\pi^{\alpha\beta}(\vx',t')\ra\,,
\ee
which can be calculated perturbatively. The viscous stress tensor can be
extracted from the energy momentum tensor using the formula
\be
\pi_{\alpha\beta}(x)=(\De_{\alpha}^{\rho}\De_{\beta}^{\sg}
-\frac{1}{3}\De_{\alpha\beta}\De^{\rho\sg})T_{\rho\sg}(x)
\ee
where $T_\mn=-g_\mn\cl+\frac{\del
\cl}{\del(\del^\mu\vph)}\cdot{\del_\nu\vec\phi}$  in which 
$\vph (x)$ denotes the pion triplet. We take the lowest order chiral
Lagrangian given by~\cite{Gasser}
\ba
\cl&=&\frac{1}{2}\del_\mu\vph\cdot\del^\mu\vph-\frac{1}{2}m_\pi^2\vph\cdot\vph+\cl_{int}\nonumber\\
\cl_{int}
&=&-\frac{1}{6f_\pi^2}[\del_\mu\vph\cdot\del^\mu\vph\,\vph\cdot\vph-
\vph\cdot\del_\mu\vph\,\vph\cdot\del^\mu\vph]+\frac{m_\pi^2}{4!~f_\pi^2}(\vph\cdot\vph)^2~.
\ea    
The time-ordered correlator, to leading order, is then given by Wick 
contractions of pion fields in $\pi_{\alpha\beta}$ which is obtained as 
\be
\pi_{\alpha\beta}(x)=(\De_{\alpha}^{\rho}\De_{\beta}^{\sg}
-\frac{1}{3}\De_{\alpha\beta}\De^{\rho\sg})\del_\rho\vph (x)\cdot\del_\sg\vph
(x)~.
\label{piab}
\ee  
In the so-called skeleton expansion, these contractions are expressed in terms of 
complete propagators (see Fig.~2) to get,
\be
\Pi_{11}(q)=i\int \frac{d^4l}{(2\pi)^4}N(l,q)D_{11}(l)D_{11}(l-q)\,,
\label{pi11}
\ee
where $D_{11}$ is given by eq.~(\ref{d11matrix}) and 
$N(l,q)$ is determined by the derivatives acting on the pion fields,
\be
N(l,q)=-g_\pi\,[\vl^2(\vl-\vq)^2+\frac{1}{3}\{\vl\cdot(\vl-\vq)\}^2]
\label{nfac}
\ee
where the pion isospin degeneracy factor $g_\pi=3$.

To work out the $l_0$ integral in eq.~(\ref{pi11}), it is more convenient to use $\Lm_{11}$
given by eq.~(\ref{Lm1}) than by eq.~(\ref{Lm2}). Closing the contour in the upper or lower 
half $l_0$-plane we get
\be
\Pi_{11}(q)=\int \frac{d^3l}{(2\pi)^3}N(\vl,\vq)\int\frac{dk_0'}{2\pi}
\rho(k_0',\vl)\frac{dk_0''}{2\pi}\rho(k_0'',\vl-\vq) K(q_0,k_0',k_0'')\,,
\ee
where 
\be
K=\frac{\{1+f(k_0')\}f(k_0'')}{q_0-(k_0'-k_0'')+i\eta}-
\frac{f(k_0')\{1+f(k_0'')\}}{q_0-(k_0'-k_0'')-i\eta}\,.
\ee
The imaginary part of $\Pi_{11}$ arises from the factor $K$,
\ba
{\rm Im}K&=&-\pi\left[\{1+f(k_0')\}f(k_0'')+f(k_0')\{1+f(k_0'')\}\right]
\de(q_0-(k_0'-k_0''))\nonumber\\
&=&-\pi\coth(\beta q_0/2)\{f(k_0'')-f(k_0')\}\de(q_0-(k_0'-k_0''))\,,
\ea
while its real part is given by the principal value integrals.

Having obtained the real and imaginary parts of $\Pi_{11}(q)$, we use
relations similar to eq.~(\ref{tanh}) to build the $11$-element of the
diagonalised $\Pi$ matrix,
\be
\ov\Pi=\int \frac{d^3l}{(2\pi)^3}N(\vl,\vq)\int \frac{dk_0'}{2\pi}\rho(k_0',\vl)
\int \frac{dk_0''}{2\pi}\rho(k_0'',\vl-\vq)
\frac{\{1+f(k_0')\}f(k_0'')-f(k_0')\{1+f(k_0'')\}}
{q_0-(k_0'-k_0'')+i\eta\ep(q_0)}\,.
\label{pibar}
\ee
Finally $\ov\Pi$ can be continued to $\Pi_{11}^R$ by a relation similar to
eq.~(2.30), 
\be
\Pi^R_{11}=\int\frac{d^3l}{(2\pi)^3}N(\vl,\vq)\int \frac{dk_0'}{2\pi}\rho(k_0',\vl)
\frac{dk_0''}{2\pi}\rho(k_0'',\vl-\vq)
\frac{\{1+f(k_0')\}f(k_0'')-f(k_0')\{1+f(k_0'')\}}
{q_0-(k_0'-k_0'')+i\eta}\,.
\label{pi11r2}
\ee
Note that in eqs.~(\ref{pibar},\ref{pi11r2}) we retain the $f(k_0')f(k_0'')$ terms in the
numerator to put it in a more convenient form. Change the signs of $k_0'$ and $k_0''$ 
in the first and second term respectively. Noting relations like $1+f(-k_0)=-f(k_0)$ 
and $\rho(-k_0)=-\rho(k_0)$ we get
\be
\Pi^R_{11}(q)=\int\frac{d^3l}{(2\pi)^3}N(\vl,\vq)\int\frac{dk_0'}{2\pi}\frac{dk_0''}{2\pi}
\rho(k_0',\vl)\rho(k_0'',\vl-\vq)f(k_0')f(k_0'')W(q_0,k_0'+k_0'')\,,
\label{pi11r3}
\ee
where
\be
W=\frac{1}{q_0+k_0'+k_0''+i\eta}-\frac{1}{q_0-(k_0'+k_0'')+i\eta}\,.
\ee

Returning to the expression (\ref{eta}) for $\eta$, we now get the three-dimensional
spatial integral of the retarded correlation function by setting $\vq =0$ in
eq.~(\ref{pi11r1}) and Fourier inverting with respect to $q_0$,
\be
i\int d^3x'\,\th(-t')\la [\pi^{\alpha\beta}(\vec 0,0),\pi_{\alpha\beta}(\vx',t')]\ra
=-\int dq_0\,e^{iq_0t'}\Pi^R_{11}(q_0,\vq=0)\,.
\label{inv}
\ee
This completes our use of the real time formulation to get the required result.
The integrals appearing in the expression for $\eta$ have been evaluated in Refs. 
\cite{Hosoya,Lang}, which we describe below for completeness.

As shown in Ref. \cite{Hosoya}, the integral over $t_1,\,t'$ and $q_0$ in
eqs.~(\ref{eta}) and (\ref{inv}) may be carried out trivially to give
\be
\eta=\left.\frac{i}{10}\frac{d}{dq_0}\Pi^R_{11}(q_0)\right|_{q_0=0}\,.
\ee
The $q_0$ dependence of $\Pi^R_{11}$ is contained entirely in $W$,
\be
\left.\frac{d}{dq_0}W(q_0)\right|_{q_0=0}=-\frac{1}{(k_0'+k_0''-i\eta)^2}+
\frac{1}{(k_0'+k_0''+i\eta)^2}=2\pi i\de'(k_0'+k_0'')\,.
\ee
Changing the integration variables in eq.~(\ref{pi11r3}) from $k_0'$, $k_0''$ to 
$\ov k_0=k_0'+k_0''$ and $k_0=\frac{1}{2}(k_0'-k_0'')$ we get
\be
\eta=\int \frac{d^3l}{(2\pi)^3}N(\vl)\int \frac{dk_0}{(2\pi)^2}F(k_0,\vl)\,,
\label{eta2}
\ee
where
\be
F(k_0,\vl)=\left.\frac{d}{d\ov k_0}\left\{\rho\!
\left(\frac{\ov k_0}{2}+k_0,\vl\right)\rho\!\left(\frac{\ov k_0}{2}-k_0,\vl\right)
f\!\left(\frac{\ov k_0}{2}+k_0\right)f\!\left(\frac{\ov
k_0}{2}-k_0\right)\right\}\right|_{\ov k_0=0}\,.
\ee

It turns out that the integral over $k_0$ becomes undefined, if we try to evaluate
$F(k_0)$ with the free spectral function $\rho_0(k)$ given by eq.~(\ref{freesp}).
As pointed out in Ref.~\cite{Hosoya}, we have to take the spectral function
for the complete propagator that includes the self-energy of the pion,
leading to its finite width $\Gm$ in the medium,
\be
\rho(k_0,\vl)=\frac{1}{i}\left[\frac{1}{(k_0-i\Gm)^2-\om^2}
-\frac{1}{(k_0+i\Gm)^2-\om^2}\right]\,,~~~~~~~~~~\om=\sqrt{\vl^2+m_\pi^2}\,.
\ee
Note that this form of the spectral function trivially follows on replacing
$i\eta$ (where $\eta\to 0^+$) with $i\Gm$ in the free spectral function (\ref{freesp}) which can
be written as
\be
\rho_0(k_0,\vl)=\frac{1}{i}\left[\frac{1}{(k_0-i\eta)^2-\om^2}
-\frac{1}{(k_0+i\eta)^2-\om^2}\right]\,.
\ee

Then $F(k_0,\vl)$ becomes
\be
F=-8\frac{k_0^2e^{\bet k_0}}{(e^{\bet k_0}-1)^2}
\frac{\bet\Gm^2}{\{(k_0-i\Gm)^2-\om^2\}^2\{(k_0+i\Gm)^2-\om^2\}^2}\,,
\ee
having double poles at $k_0=2\pi in/\bet$ for $n=\pm 1, \pm 2, \cdots$ and
also at $k_0=\pm\om\pm i\Gm$. The integral over $k_0$ may now be evaluated by
closing the contour in the upper/lower half-plane to get
\be
\int^{+\infty}_{-\infty}\frac{dk_0}{(2\pi)^2}
F(k_0,\vl)=-\frac{1}{8\pi}\frac{\bet}{\om^2\Gm}n(\om)\{1+n(\om)\}\,,
\ee
where we retain only the leading (singular) term for small $\Gm$. In this
approximation eq.~(\ref{eta2}) gives
\be
\eta=\frac{g_\pi\beta}{30\pi^2}\int_0^\infty dl\,
\frac{l^6}{\om^2\Gm}{n(\om)\{1+n(\om)\}}\,.
\label{eta_lrt}
\ee

Proceeding analogously as above, the lowest order contribution to
the bulk viscosity can be obtained as~\cite{Hosoya}
\be
\zeta=\frac{g_\pi\beta}{4\pi^2}\int_0^\infty dl\,
\frac{l^2(l^2/3-c_s^2\om^2)^2}{\om^2\Gm}n(\om)\{1+n(\om)\}\,.
\label{zeta_lrt}
\ee

The width $\Gm (l)$ at different temperatures is known \cite{Goity} from chiral 
perturbation theory. The quantity $\Gm$ can also be interpreted as the collision
frequency, the inverse of which is the relaxation time $\tau$. For collisions of the form
$\pi(l)+\pi(k)\to\pi(l')+\pi(k')$ this is given by(see e.g.~\cite{Sukanya1})
\be
\Gm (l)=\tau^{-1}(l)=\int \frac{k^2\, dk}{2\pi^2\om_k}
\frac{\sqrt{s(s-4m_\pi^2)}}{2\om_l}n(\om_k)(1+n(\om_{l'}))(1+n(\om_{k'}))\frac{1}{2}\int
d\Omega\frac{d\sigma}{d\Omega}
\label{om_relax}
\ee
where  $\frac{d\sigma}{d\Omega}$ is the
$\pi\pi$ cross-section. Note that the lowest order
formulae for the shear and bulk viscosities obtained above in the linear
response approach {\em coincide} with the expressions which result from 
solving the transport equation in the relaxation-time approximation.

\section{Viscous coefficients in the kinetic theory approach}
\setcounter{equation}{0}
\renewcommand{\theequation}{3.\arabic{equation}}

The kinetic theory approach is suitable for studying transport properties
of dilute systems. Here one assumes that the system is characterized by a
distribution function which gives the phase space probability density of the
particles making up the fluid. Except during collisions, these (on-shell)
particles are assumed to propagate classically with  well defined position,
momenta and energy. It is possible to obtain the non-equilibrium distribution
function by solving the transport 
equation in the hydrodynamic regime by expanding the distribution function in a
local equilibrium part along with non-equilibrium corrections. This 
expansion in terms of gradients of the velocity field is used
to linearize the transport equation. The coefficients of expansion which are related
to the transport coefficients, satisfy linear integral equations.  
The standard method of solution involves the use of polynomial functions to 
reduce these integral equations to algebraic ones.

\subsection{Transport coefficients at first Chapman-Enskog order}
The evolution of the phase space distribution of the pions
is governed by the (transport) equation
\be
p^\mu\partial_\mu f(x,p)=C[f]
\label{treq}
\ee
where $C[f]$ is the collision integral. For binary elastic 
collisions $p+k\to p'+k'$ which we consider,
this is given by~\cite{Davesne}
\ba
C[f]&=&\int d\Gamma_k\ d\Gamma_{p'}\ d\Gamma_{k'}[f(x,p')f(x,k') \{1+f(x,p)\}
\{1+f(x,k)\}\nonumber\\
&&-f(x,p)f(x,k)\{1+f(x,p')\}\{1+f(x,k')\}]\ W
\ea
where the interaction rate,
\[
W=\frac{s}{2}\ \frac{d\sigma}{d\Omega}(2\pi)^6\delta^4(p+k-p'-k')
\]
and $d\Gamma_q=\frac{d^3q}{(2\pi)^3q_0}$. 
The $1/2$ factor comes from the indistinguishability of the initial state pions.

For small deviation from local equilibrium we write, in the first Chapman-Enskog
approximation
\be
f(x,p)=f^{(0)}(x,p)+\de f(x,p),~~~\de f(x,p)=f^{(0)}(x,p)[1+f^{(0)}(x,p)]\phi(x,p)
\label{ff}
\ee
where the equilibrium distribution function (in the new notation) is given by
\be
f^{(0)}(x,p)=\left[e^{\frac{p^{\mu}u_{\mu}(x)-\mu_\pi(x)}{T(x)}}-1\right]^{-1},
\label{f0}
\ee 
with $T(x)$, $u_\mu(x)$ and $\mu_\pi(x)$ representing the local temperature, 
flow velocity and pion chemical potential respectively.
Putting (\ref{ff}) in (\ref{treq}) 
the deviation function $\phi(x,p)$ is seen to satisfy
\be
p^\mu\partial_\mu f^{(0)}(x,p)=-\cl[\phi]
\label{treq2}
\ee
where the linearized collision term
\ba
\cl[\phi]=&&f^{(0)}(x,p)\int d\Gamma_k\ d\Gamma_{p'}\ d\Gamma_{k'}f^{(0)}(x,k)
\{1+f^{(0)}(x,p')\}\{1+f^{(0)}(x,k')\}\nonumber\\
&&[\phi(x,p)+\phi(x,k)-\phi(x,p')-\phi(x,k')]\ W~.
\ea 
Using the form of $f^{(0)}(x,p)$ as given in (\ref{f0}) on the left side of
(\ref{treq2}) and eliminating time  derivatives with the help of equilibrium 
thermodynamic laws we arrive at~\cite{Sukanya2}
\be
[Q\partial_{\nu}u^{\nu}+p_{\mu}\De^{\mu\nu}(p\cdot u- h)
(T^{-1}\partial_{\nu}T-Du_{\nu})-
\langle p_{\mu}p_{\nu} \rangle \langle \del^{\mu} u^{\nu}
\rangle]f^{(0)}(1+f^{(0)})
=-T\cl[\phi]
\label{treq3}
\ee
where $D=u^\mu \partial_\mu$, $\nabla_\mu =\Delta_{\mu\nu} \partial^\nu$,
$\Delta_{\mu\nu}=g_\mn-u_\mu u_\nu$ and $\la\cdot\ra$ indicates a space-like  
symmetric and traceless combination. In this equation
\be
Q=-\frac{1}{3}m_\pi^2+(p\cdot u)^2\{\frac{4}{3}-\gamma'
\}+p\cdot u\{(\gamma''-1) h
-\gamma'''T \}
\ee
where
\be
\gamma'=\frac{(S_{2}^{0}/S_{2}^{1})^2-(S_{3}^{0}/S_{2}^{1})^2+4z^{-1}S_{2}^{0}S_{3}^{1}/(S_{2}^{1})^2+z^{-1}S_{3}^{0}/S_{2}^{1}}
{(S_{2}^{0}/S_{2}^{1})^2-(S_{3}^{0}/S_{2}^{1})^2+3z^{-1}S_{2}^{0}S_{3}^{1}/(S_{2}^{1})^2+2z^{-1}S_{3}^{0}/S_{2}^{1}-z^{-2}}
\ee
\be
\gamma''=1+\frac{z^{-2}}
{(S_{2}^{0}/S_{2}^{1})^2-(S_{3}^{0}/S_{2}^{1})^2+3z^{-1}S_{2}^{0}S_{3}^{1}/(S_{2}^{1})^2+2z^{-1}S_{3}^{0}/S_{2}^{1}-z^{-2}}
\ee
\be
\gamma'''=\frac{S_{2}^{0}/S_{2}^{1}+5z^{-1}S_{3}^{1}/S_{2}^{1}-S_{3}^{0}S_{3}^{1}/(S_{2}^{1})^2}
{(S_{2}^{0}/S_{2}^{1})^2-(S_{3}^{0}/S_{2}^{1})^2+3z^{-1}S_{2}^{0}S_{3}^{1}/(S_{2}^{1})^2+2z^{-1}S_{3}^{0}/S_{2}^{1}-z^{-2}}
\ee
with $z=m_{\pi}/T$ and $h=m_\pi S_{3}^{1}/S_{2}^{1}$.  
The functions $S_n^\alpha(z)$ are integrals over Bose functions~\cite{Sukanya2} and are defined
as $S_n^\alpha(z)=\sum_{k=1}^\infty e^{k\mu_\pi/T} k^{-\alpha} K_n(kz)$, 
$K_n(x)$ denoting the modified Bessel function of order $n$. 
The left hand side of (\ref{treq2}) is thus expressed in terms of thermodynamic forces
with different tensorial ranks. In order to be a solution of this equation
$\phi$ must also be a linear combination of the corresponding
thermodynamic forces.
It is typical to take $\phi$ as
\be
\phi=A\partial\cdot u+B_{\mu}\nabla^{\mu\nu}(T^{-1}\partial_{\nu}T
-Du_{\nu})-C_{\mu\nu}\langle \partial^{\mu} u^{\nu} \rangle
\label{phi}
\ee
which on 
substitution into (\ref{treq3}) and comparing coefficients of the
(independent) thermodynamic forces on both sides, yields the set of equations
\be
\cl[A]=-Q f^{(0)}(p)\{1+f^{(0)}(p)\}/T
\label{AA}
\ee
\be                             
\cl[C_\mn]=-\la p_\mu p_\nu\ra f^{(0)}(p)\{1+f^{(0)}(p)\}/T
\label{CC}
\ee  
ignoring the equation for $B_\mu$ which is related to thermal conductivity.
These integral equations are to be solved to get the coefficients
$A$ and $C_\mn$. It now remains to link these to the viscous coefficients
$\zeta$ and $\eta$. This is achieved by means of the dissipative part of the 
energy-momentum tensor resulting from the use of the non-equilibrium 
distribution function (\ref{ff}) in
\be
T^\mn=\int d\Gamma_p\ p^\mu p^\nu f(p)=T^{\mn (0)}+\De T^{\mn}
\ee
where
\be
\Delta T^{\mu\nu}=\int d\Gamma_p f^{(0)} (1+f^{(0)}) C_{\alpha\beta} 
\langle p^{\alpha}p^{\beta} 
\rangle \langle \partial^{\mu} u^{\nu} \rangle
+\int d\Gamma_p f^{(0)} (1+f^{(0)}
)QA\Delta^{\mu\nu}\partial_{\sigma}u^{\sigma}~.
\ee
Again, for a small deviation $\phi(x,p)$, close to equilibrium, so that
only first order derivatives contribute, the dissipative tensor can be
generally expressed in the form~\cite{Purnendu,Polak}
\be
\Delta T^{\mu\nu}=-2\eta\langle \partial^{\mu} u^{\nu} \rangle
-\zeta\Delta^{\mu\nu}\partial_{\sigma}u^{\sigma}~.
\ee
Comparing,
we obtain the expressions of shear and bulk viscosity,
\be
\eta=-\frac{1}{10}\int d\Gamma_p\ C_\mn\la p^\mu p^\nu \ra
f^{(0)}(p)\{1+f^{(0)}(p)\}
\ee
and
\be
\zeta=-\int d\Gamma_p \ QA f^{(0)}(p)\{1+f^{(0)}(p)\}~.
\ee
The coefficients $A$ and $C_{\mu\nu}$ are perturbatively obtained from
(\ref{AA}) and (\ref{CC})
by expanding in terms of orthogonal polynomials which reduces the integral
equations to algebraic ones. After a tedious calculation
using the Laguerre polynomial of 1/2 integral order, the first 
approximation to the shear and bulk viscosity come out as
\be
\eta=\frac{T}{10}\ \frac{\gamma_0^2}{c_{00}}
\label{eta_KT}
\ee
and
\be
\zeta=T\frac{\alpha_{2}^{2}}{a_{22}}
\label{zeta}
\ee
where
\ba
\gamma_0&=&-10\frac{S_3^{2}(z)}{S_2^{1}(z)}~,\nonumber\\
c_{00}&=&16\{I_1(z)+I_2(z)+\frac{1}{3}I_3(z)\}~,
\ea
and
\ba
\alpha_{2}&=&\frac{z^3}{2} [ \frac{1}{3}(\frac{S_{3}^0}{S_{2}^{1}}-z^{-1})
+(\frac{S_{2}^{0}}{S_{2}^{1}}+\frac{3}{z}\frac{S_{3}^{1}}{S_{2}^{1}})
\{(1-\gamma'')\frac{S_{3}^{1}}{S_{2}^{1}}+\gamma'''z^{-1}\}
\nonumber\\
&-&(\frac{4}{3}-\gamma')\{ \frac{S_{3}^{0}}
{S_{2}^{1}}+15z^{-2}\frac{S_{3}^{2}}{S_{2}^{1}}+2z^{-1}\}]~,
\nonumber\\a_{22}&=&2z^2I_{3}(z)~.
\ea

The integrals $I_\alpha(z)$ are given by
\ba
I_\alpha(z)&=&\frac{z^4}{[S_2^{1}(z)]^2} \ e^{(-2\mu_\pi/T)}\int_0^\infty d\psi\ \cosh^3\psi
\sinh^7\psi\int_0^\pi
d\Theta\sin\Theta\frac{1}{2}\frac{d\sigma}{d\Omega}(\psi,\Theta)\int_0^{2\pi}
d\phi\nonumber\\&&\int_0^\infty d\chi \sinh^{2\alpha} \chi
\int_0^\pi d\theta\sin\theta\frac{e^{2z\cosh\psi\cosh\chi}}
{(e^E-1)(e^F-1)(e^G-1)(e^H-1)}\ M_\alpha(\theta,\Theta)
\label{c00_bose}
\ea
where
$\mu_\pi$ is the chemical potential of pions. The exponents in the Bose functions
are given by
\ba
E&=&z(\cosh\psi\cosh\chi-\sinh\psi\sinh\chi\cos\theta)-\mu_\pi/T\nonumber\\
F&=&z(\cosh\psi\cosh\chi-\sinh\psi\sinh\chi\cos\theta')-\mu_\pi/T\nonumber\\
G&=&E+2z\sinh\psi\sinh\chi\cos\theta\nonumber\\
H&=&F+2z\sinh\psi\sinh\chi\cos\theta'~,
\ea

and the functions $M_\alpha(\theta,\Theta)$ represent
\ba
M_1(\theta,\Theta)&=&1-\cos^2\Theta~,\nonumber\\M_2(\theta,\Theta)&=&\cos^2\theta+\cos^2\theta'
-2\cos\theta\cos\theta'\cos\Theta~,\nonumber\\
M_3(\theta,\Theta)&=&[\cos^2\theta-\cos^2\theta']^2~.
\ea
The relative angle $\theta^\prime$ is defined by,
\(
\cos\theta'=\cos\theta\cos\Theta-\sin\theta\sin\Theta\cos\phi~.
\)

Note that the differential cross-section which appears in the denominator
is the dynamical input in the expressions for $\eta$ and $\zeta$. It is this
quantity we turn to in the next section.

\subsection{The $\pi\pi$ cross-section with medium effects} 

\begin{figure}
\includegraphics[scale=0.3]{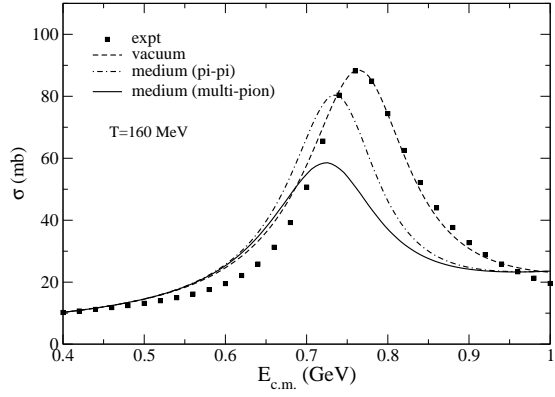}
\caption{The $\pi\pi$ cross-section as a function of centre of mass energy. The
dotted line indicates the cross-section obtained using eq.~(\ref{amp}) which
agrees well with the experimental values (eq.~(\ref{expt})) shown by filled
squares. The dashed
and solid lines depict the in-medium cross-section for $\pi\pi$ and $\pi
h(h=\pi,\om,h_1,a_1)$ 
loops respectively in the $\rho$ self-energy evaluated at $T$=160 MeV.}
\label{sigmafig}
\end{figure}

The strong interaction dynamics of the pions enters the collision integrals
through the cross-section. In Fig.~\ref{sigmafig} we show the $\pi\pi$
cross-section as a function of the centre of mass energy of scattering. The
different curves are explained below.
The filled squares referred to as experiment is a widely used resonance saturation
parametrization~\cite{Bertsch1,Prakash} of isoscalar and isovector phase shifts obtained from various
empirical data involving the $\pi\pi$ system. The isospin averaged 
differential cross-section
is given by
\be
\frac{d\sigma(s)}{d\Omega}=\frac{4}{q_{cm}^2}\left[\frac{1}{9}
\sin^2\de^0_0+\frac{5}{9}\sin^2\de^2_0+\frac{1}{3}\cdot
9\sin^2\de_1^1\cos^2\theta\right]
\label{expt}
\ee
where
\ba
\de^0_0&=&\frac{\pi}{2}+\arctan\left(\frac{E-m_\sigma}{\Gamma_\sigma/2}\right)\nonumber\\
\de_1^1&=&\frac{\pi}{2}+\arctan\left(\frac{E-m_\rho}{\Gamma_\rho/2}\right)\nonumber\\
\de^2_0&=&-0.12p/m_\pi~.
\ea
The widths are given by $\Gamma_\sigma=2.06p$ and 
$\Gamma_\rho=0.095p\left(\frac{p/m_\pi}{1+(p/m_\rho)^2}\right)^2$
with $m_\sigma=5.8m_\pi$ and $m_\rho=5.53m_\pi$~.

To get a handle on the dynamics we now evaluate the $\pi\pi$ cross-section
involving $\rho$ and $\sigma$ meson exchange processes using the interaction
Lagrangian
\be
\cl=g_\rho\vec\rho^\mu\cdot\vec \pi\times\del_\mu\vec \pi+\frac{1}{2}g_\sigma
m_\sigma\vec \pi\cdot\vec\pi\sigma
\ee
where $g_\rho=6.05$ and $g_\sigma=2.5$.
In the matrix elements corresponding to $s$-channel $\rho$ and $\sigma$
exchange diagrams which appear for total isospin $I=1$ and 0 respectively, we
introduce a decay width in the corresponding propagator. We get~\cite{Sukanya1},
\ba
\cm_{I=0}&=&
g_\sigma^2 m_\sigma^2\left[\frac{3}{s-m_\sg^2+im_\sg\Gm_\sg}+\frac{1}{t-m_\sg^2}
+\frac{1}{u-m_\sg^2}\right]
+2g_\rho^2\left[\frac{s-u}{t-m_\rho^2}+\frac{s-t}{u-m_\rho^2}\right]
\nonumber\\
\cm_{I=1}&=&g_\sigma^2 m_\sigma^2\left[\frac{1}{t-m_\sg^2}-\frac{1}{u-m_\sg^2}\right]
+g_\rho^2\left[\frac{2(t-u)}{s-m_\rho^2+im_\rho\Gamma_\rho}+
\frac{t-s}{u-m_\rho^2}-\frac{u-s}{t-m_\rho^2}\right]
\nonumber\\
\cm_{I=2}&=&g_\sigma^2 m_\sigma^2\left[\frac{1}{t-m_\sg^2}+\frac{1}{u-m_\sg^2}\right]
+g_\rho^2\left[\frac{u-s}{t-m_\rho^2}+\frac{t-s}{u-m_\rho^2}\right]~.
\label{amp}
\ea
The differential cross-section is then obtained from 
$\frac{d\sigma}{d\Omega}=\overline{|\cm|^2}/64\pi^2 s$ 
where the isospin averaged amplitude is given by
$\overline{|\cm|^2}=\frac{1}{9}\sum(2I+1)\overline{|\cm_I|^2}$.
 
The integrated cross-section, 
after ignoring the $I=2$ contribution is shown by 
the dotted line (indicated by 'vacuum') in Fig.~\ref{sigmafig} and 
is seen to agree 
reasonably well with the experimental cross-section 
up to a centre of mass energy of about 1 GeV beyond which the theoretical
estimate gives higher values. We hence use the experimental cross-section beyond
this energy.

\begin{figure}
\includegraphics[scale=0.5]{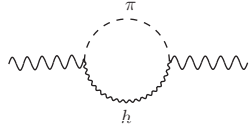}
\caption{$\pi h$ self-energy diagrams where $h$ stands for
$\pi,\om,h_1,a_1$ mesons.}
\label{selfdiag}
\end{figure}

After this normalisation to data, we now turn to the in-medium cross-section by introducing
the effective propagator for the $\rho$ in the above expressions for the matrix
elements. This is obtained in terms of the self-energy by solving the Dyson
equation and is given by
\be
D_\mn=D^{(0)}_\mn+D^{(0)}_{\mu\sigma}\Pi^{\sigma\lambda}D_{\lambda\nu}
\label{eq:dyson}
\ee
where $D^{(0)}_\mn$ is the vacuum propagator for the $\rho$ meson
and $\Pi^{\sigma\lambda}$ is the self energy function obtained
from one-loop diagrams shown in Fig.~\ref{selfdiag}.
The standard procedure~\cite{Bellac} to solve this equation in the medium is to 
decompose the self-energy into transverse and longitudinal components.
For the case at hand the difference between these components 
is found to be small and is hence ignored.
We work with the polarization averaged self-energy function defined as
\be
\Pi=\frac{1}{3}(2\Pi^T+q^2\Pi^L)
\ee
where
\be
\Pi^T=-\frac{1}{2}(\Pi_\mu^\mu +\frac{q^2}{\bar q^2}\Pi_{00}),~~~~
\Pi^L=\frac{1}{\bar q^2}\Pi_{00} , ~~~\Pi_{00}\equiv u^\mu u^\nu \Pi_{\mn}~.
\label{pitpil}
\ee
The in-medium propagator is then written as
\be
D_\mn(q_0,\vec q)=\frac{-g_\mn+q_\mu q_\nu/q^2}{q^2-m_\rho^2-{\rm Re}\Pi(q_0,\vec q)+
i{\rm Im}\Pi(q_0,\vec q)}~.
\label{medprop}
\ee
The scattering, decay and regeneration processes which cause a gain or loss of
$\rho$ mesons in the medium are responsible for the imaginary part of its 
self-energy. 
The real part on the other hand modifies the position
of the pole of the spectral function.

As discussed in Sec.~IIA, in the real-time formulation of thermal field theory the self-energy
assumes a 2$\times$2 matrix structure of which the 11-component is given by 
\be
\Pi_{\mn}^{11}(q)=i\int\frac{d^4k}{(2\pi)^4}N_{\mn}(q,k)D_\pi ^{11}(k)D_h^{11}(q-k)
\ee
where $D^{11}$ is the 11-component of the scalar propagator given by
$D^{11}(k)=\De(k)+2\pi if^{(0)}(k)\de(k^2-m^2)$. It turns out that the self-energy
function mentioned above can be obtained in terms of the 11-component through the
relations~\cite{Bellac,Mallik_RT}
\ba
{\rm Re}\,\Pi_{\mn}&=&{\rm Re}\,\Pi_{\mn}^{11}\nonumber\\
{\rm Im}\Pi_{\mn}&=&\epsilon(q_0)\tanh(\beta q_0/2){\rm Im}\,\Pi_{\mn}^{11}~.
\ea
Tensor structures associated with the two  
vertices and the vector propagator are included in $N_{\mn}$ and are
available in~\cite{Ghosh1} where the interactions were
taken from chiral perturbation theory.
It is easy to perform the integral over $k_0$ using suitable contours to obtain
\ba
{\Pi}^{\mn}(q_0,\vq)&=&\int\frac{d^3k}{(2\pi)^3}\frac{1}
{4\om_{\pi}\om_{h}}\left[\frac{(1+f^{(0)}(\omp))N^{\mn}_1+f^{(0)}(\omh)N^{\mn}_3}
{q_0 -\om_{\pi}-\om_{h}+i\eta\ep(q_0)}
+\frac{-f^{(0)}(\omp)N^{\mn}_1+f^{(0)}(\omh)N^{\mn}_4}
{q_0-\om_{\pi}+\om_{h}+i\eta\ep(q_0)} 
\right.\nonumber\\
&&+\left. \frac{f^{(0)}(\omp)N^{\mn}_2 -f^{(0)}(\omh)N^{\mn}_3}
{q_0 +\om_{\pi}-\om_{h}+i\eta\ep(q_0)} 
+\frac{-f^{(0)}(\omp)N^{\mn}_2 -(1+f^{(0)}(\omh))N^{\mn}_4}
{q_0 +\om_{\pi}+\om_{h}+i\eta\ep(q_0)}\right]
\label{MM_rho}
\ea
where $f^{(0)}(\om)=\frac{1}{e^{(\om-\mu_\pi)/T}-1}$ is the Bose distribution
function with arguments $\om_\pi=\sqrt{\vk^2+m_\pi^2}$ and
$\om_h=\sqrt{(\vq-\vk)^2+m_h^2}$. Note that this expression is a generalized
form for the in-medium self-energy obtained by Weldon~\cite{Weldon}.
The subscript $i(=1,..4)$ on $N^{\mn}$ in (\ref{MM_rho}) correspond to its values for 
$k_0=\om_\pi,-\om_\pi,q_0-\om_h,q_0+\om_h$ respectively. It is easy to read off
the real and imaginary parts from (\ref{MM_rho}). The angular integration can be
carried out using the $\de$-functions in each of the four terms in the imaginary
part which define the kinematically allowed regions in $q_0$ and $\vq$ where
scattering, decay and regeneration processes occur in the medium leading to the
loss or gain of $\rho$ mesons~\cite{Ghosh1}. 
The vector mesons $\omega$, $h_1$ and $a_1$ which appear in the loop
have negative
$G$-parity and have substantial $3\pi$ and $\rho\pi$ decay widths~\cite{PDG}. The
(polarization averaged) self-energies containing these unstable particles in the loop graphs have thus been folded 
with their spectral functions,
\be
\Pi(q,m_h)= \frac{1}{N_h}\int^{(m_h+2\Gm_h)^2}_{(m_h-2\Gm_h)^2}dM^2\frac{1}{\pi} 
{\rm Im} \left[\frac{1}{M^2-m_h^2 + iM\Gm_h(M) } \right] \Pi(q,M) 
\ee
with $N_h=\displaystyle\int^{(m_h+2\Gm_h)^2}_{(m_h-2\Gm_h)^2}
dM^2\frac{1}{\pi} {\rm Im}\left[\frac{1}{M^2-m_h^2 + iM\Gm_h(M)} \right]$.
The contributions from the
loops with heavy mesons may then be considered as a multi-pion contribution to the
$\rho$ self-energy. 

The in-medium cross-section is now obtained by using the full $\rho$-propagator
(\ref{medprop}) in place of the usual vacuum propagator $D_\mn^{(0)}$ in the 
scattering amplitudes. The long dashed line in Fig.~\ref{sigmafig} shows a 
suppression of the peak when only the $\pi\pi$ loop is considered. This
effect is magnified when the $\pi h$ loops 
(solid line indicated by multi-pion) 
are taken into account 
and is also accompanied by a small shift in the peak position.
Extension to the case of finite baryon density can be done 
using the spectral function computed in~\cite{Ghosh2} where an extensive
list of baryon (and
anti-baryon) loops are considered along with the mesons.
A similar modification of the $\pi\pi$ cross-section for a hot and dense
system was seen also in~\cite{Bertsch2}.
\begin{figure}
\includegraphics[scale=0.3]{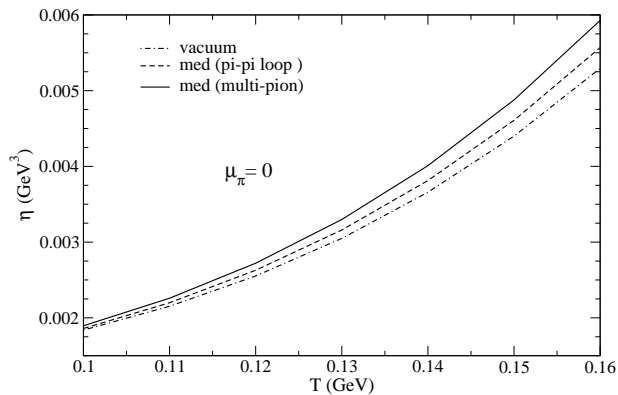}
\caption{The shear viscosity as a function of temperature in the
Chapman-Enskog approximation.  The dash-dotted line indicates use of the 
vacuum cross-section and the dashed and solid lines 
correspond to in-medium cross-section for the $\pi\pi$ and multi-pion cases 
respectively.}
\label{CEfig}
\end{figure}

\begin{figure}
\includegraphics[scale=0.3]{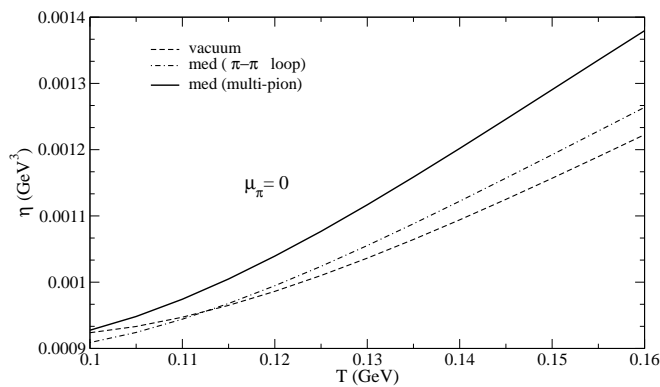}
\caption{The shear viscosity as a function of temperature in the
relaxation time approximation. The dash-dotted and solid lines correspond to the
use of in-medium cross-sections in eq.~(\ref{om_relax}) for $\pi\pi$ and
multi-pion loops respectively. The dashed line represents the vacuum case.} 
\label{relaxfig}
\end{figure}

We plot $\eta$ versus $T$ in Fig.~\ref{CEfig} obtained in the Chapman-Enskog
approximation showing the effect of the in-medium $\rho$ propagation in
the pion gas~\cite{Sukanya1}. We observe  $\sim 10\%$ change at $T=150$ MeV due to medium effects 
compared to the vacuum when all the loops in the $\rho$ self-energy are considered.
The effect reduces with temperature to less than $5\%$ at 100 MeV. 

We noted in Sec.~II that the lowest order result for $\eta$ in the response theory framework coincides with
that obtained in the relaxation time approximation which is in fact the simplest way 
to linearize the transport equation. Here one assumes that $f(x,p)$ goes over to the equilibrium distribution
$f^{(0)}(x,p)$ as a result of collisions and this takes place over a relaxation
time $\tau(p)$ which is the inverse of the collision frequency defined in
(\ref{om_relax}). The
right hand side of eq.~(\ref{treq}) is then given by
$-E_p[f(x,p)-f^{(0)}(x,p)]/ \tau(p)$ which subsequently leads to the expressions
(\ref{eta_lrt}) and (\ref{zeta_lrt}) for the shear and bulk
viscosities~\cite{Gavin}.
In Fig.~\ref{relaxfig} we show the temperature dependence of $\eta$ in the relaxation time
approximation. The values in this case are lower than
that obtained in the Chapman-Enskog method though the effect of the medium is
larger. In addition to the fact that the expressions for the viscosities are quite different
in two approaches, the difference in the numerical values obtained in the two
cases also depends significantly on the energy dependence of the $\pi\pi$
cross-section~\cite{Wiranata}.

\begin{figure}
\includegraphics[scale=0.4]{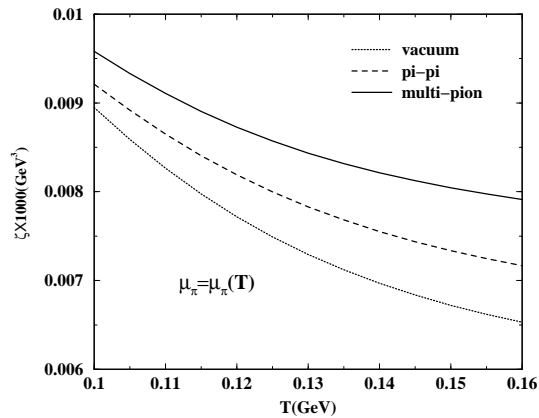}
\caption{The bulk viscosity as a function of temperature for a chemically frozen
pion gas. The dashed and solid lines correspond to the
use of in-medium cross-sections in eq.~(\ref{om_relax}) for $\pi\pi$ and
multi-pion loops respectively. The dotted line represents the vacuum case. } 
\label{zetafig}
\end{figure}

In Fig.~\ref{zetafig} we show the numerical results for the bulk viscosity of a
pion gas as function of $T$. It is seen from an analysis of the left hand side
of the transport equation that while the shear viscosity depends on elastic
processes, bulk viscosity is sensitive to number changing processes. However
in heavy ion collision experiments matter is known to undergo early chemical
freeze-out. Number changing (inelastic)
processes having much larger relaxation times go out of equilibrium at this 
point and a temperature dependent chemical potential results for each species so
as to conserve the number corresponding to the measured particle ratios. 
We hence use a temperature dependent pion chemical potential taken 
from~\cite{Hirano} in this case. It is interesting to observe that 
$\zeta$ decreases with $T$
in contrast to $\eta$ which increases. The trend followed by $\zeta$ is similar
to the findings of~\cite{Moore}. Additional discussions concerning the temperature
dependence of viscosities for a chemically frozen pion gas are available
in~\cite{Sukanya2}.  

\section{Summary and Conclusion}

To summarize, we have calculated the shear viscosity coefficient of a pion gas
in the real time version of thermal field 
theory. It is simpler to the imaginary version in that we do not have to continue to 
imaginary time at any stage of the calculation. As an element in the theory of linear 
response, a transport coefficient is defined in terms of a retarded thermal two-point 
function of the components of the energy-momentum tensor. We derive K\"{a}llen-Lehmann
representation for any (bosonic) two-point function of both time-ordered and
retarded types to get the relation between them. Once this relation is
obtained, we can calculate the retarded function in the Feynman-Dyson
framework of the perturbation theory. 

Clearly the method is not restricted to transport coefficients. Any linear
response leads to a retarded two-point function, which can be calculated in
this way. Also quadratic response formulae have been derived in the real
time formulation \cite{Carrington}.

We have also evaluated the viscous coefficients in the kinetic theory approach to leading order in
the Chapman-Enskog expansion. Here we have incorporated an in-medium $\pi\pi$
cross-section and found a significant effect in the temperature dependence
of the shear viscosity. 

The viscous coefficients and their temperature dependence could affect
the quantitative estimates of signals of heavy ion collisions particularly
where hydrodynamic simulations are involved.
For example, it has been argued in~\cite{Dusling} that corrections to the freeze-out distribution
due to bulk viscosity can be significant. As a result the hydrodynamic 
description of the $p_T$ spectra and elliptic flow of hadrons could be improved
by including a realistic temperature dependence of the viscous coefficients.
Such an evaluation essentially requires the consideration of a multi-component
gas preferably containing nucleonic degrees of freedom so that extensions to
finite baryon chemical potential can be made. Work in this direction is in progress. 

\section*{Acknowledgement}
The author gratefully acknowledges the contribution from his collaborators
S. Mallik, S. Ghosh and S. Mitra to various topics presented here.

\end{document}